\documentclass[pra,twocolumn,superscriptaddress,amssymb]{revtex4}
\usepackage{graphicx}
\usepackage{bm}
\usepackage{amsmath,amssymb,fancybox}
\usepackage{color}

\begin{document}
\title{Measurement-only verifiable blind quantum computing with quantum input verification} 
\author{Tomoyuki Morimae}
\email{morimae@gunma-u.ac.jp}
\affiliation{ASRLD Unit, Gunma University, 1-5-1 Tenjin-cho Kiryu-shi
Gunma-ken, 376-0052, Japan}

\begin{abstract}
Verifiable blind quantum computing is a secure delegated quantum
computing where a client with a limited quantum technology 
delegates her quantum computing to a server who has 
a universal quantum computer. The client's privacy
is protected (blindness) and the correctness of the computation 
is verifiable by the 
client in spite of her limited quantum technology (verifiability).
There are mainly two types of protocols for verifiable blind quantum computing:
the protocol where the client has only to generate single-qubit states,
and the protocol where the client needs only the ability of single-qubit 
measurements.
The latter is called the measurement-only verifiable blind quantum computing.
If the input of the client's quantum computing is a quantum state
whose classical efficient description is not known to the client,
there was no way for the measurement-only client to verify the correctness
of the input. Here we introduce
a new protocol of measurement-only verifiable blind quantum
computing where the correctness of the
quantum input is also verifiable.
\end{abstract}
\maketitle  

\section{Introduction}
Blind quantum computing is a secure delegated quantum computing 
where a client (Alice) who does not have enough quantum technology
delegates her quantum computing to a server (Bob) who has a universal
quantum computer without leaking any information about her quantum computing.
By using measurement-based quantum computing~\cite{MBQC,MBQC2},
Broadbent, Fitzsimons, and Kashefi first showed that
blind quantum computing is indeed possible for a client
who can do only the single qubit state generation~\cite{BFK}. 
Since the breakthrough, many theoretical 
improvements have been obtained~\cite{MFNatComm,
CV,coherent,Lorenzo,Mantri,distillation,composable,AKLT,Sueki},
and even a proof-of-principle experiment was achieved
with photonic qubits~\cite{experiment1}.
These blind quantum computing protocols 
guarantee two properties: first, if Bob is honest, Alice can obtain
the correct result of her quantum computing (correctness).
Second, whatever Bob does, he cannot gain any information about
Alice's quantum computing (blindness)~\cite{unavoidable}.

In stead of the single-qubit state generation,
the ability of single-qubit measurements is also enough for
Alice: it was shown in Ref.~\cite{measuringAlice}
that Alice who can do only single-qubit measurements can
perform blind quantum computing.
The idea is that Bob generates a graph state and sends each qubit
one by one to Alice. If Bob is honest, he generates the correct graph state
and therefore Alice can perform the correct measurement-based
quantum computing (correctness). If Bob is malicious,
he might send a wrong state to Alice, but whatever Bob sends to
Alice, Alice's measurement angles, which contain information
about Alice's computation, cannot be transmitted to Bob
due to the no-signaling principle (blindness).
The protocol is called the measurement-only protocol,
since Alice needs only measurements.

A problem in all these protocols is the lack of the verifiability:
although the blindness guarantees that
Bob cannot learn Alice's quantum computing, he can
still deviate from the correct procedure, mess up her quantum computing,
and give Alice a completely wrong result.
Since Alice cannot perform quantum computing by herself,
she cannot check the correctness of the result by herself
unless the problem is, say, in NP,
and therefore she can accept a wrong result.
To solve the problem, verifiable blind quantum computing protocol
was introduced in Ref.~\cite{FK}, and 
some theoretical improvements were also obtained~\cite{topoveri,ElhamDI,JoeDI}.
Experimental demonstrations of the verification
were also done~\cite{experiment2,experiment3}.
The basic idea of these protocols is 
so called 
the trap technique: Alice hides some trap qubits in the register,
and any change of a trap signals Bob's malicious behavior.
By checking traps, Alice can detect any Bob's malicious behavior
with high probability.
If the computation is encoded by a quantum error detection code,
the probability that Alice is fooled by Bob can be exponentially small,
since in that case in order to change the logical state,
Bob has to touch many qubits, and it 
consequently increases the probability that 
Bob touches some traps.

Recently, a new verification protocol
that does not use the trap technique
was proposed~\cite{HM} (see also Ref.~\cite{HH}). In this protocol, 
Bob generates a graph state, and 
sends each qubit of it one by one to Alice.
Alice directly verifies
the correctness of the graph state (and therefore
the correctness of the computation) sent from Bob
by measuring stabilizer operators.
This verification technique is called the stabilizer test.
Note that the stabilizer test is useful also in quantum
interactive proof system~\cite{QMAsingle,QAMsingle,Matt,FV,Ji}.

Although computing itself is verifiable
through the stabilizer test, the input is not if it is a quantum state
whose classical efficient description is not known to Alice.
For example, let us assume that Alice receives a state $|\psi\rangle$
from Charlie, and she wants to apply a unitary $U$ on $|\psi\rangle$.
If she delegates the quantum computation to Bob
in the measurement-only style, a possible procedure is as follows.
Alice first sends $|\psi\rangle$ to Bob. Bob next entangles
$|\psi\rangle$ to the graph state.
Bob then sends each qubit of the state one by one to Alice, 
and Alice does measurement-based quantum computing on it.
If Bob is honest, Alice can realize $U|\psi\rangle$.
Furthermore,
as is shown in Refs.~\cite{QMAsingle}, Alice can verify the
correctness of the graph state by using the stabilizer test
even if some states are coupled to the graph state.
However, in the procedure, the correctness of the input state is not 
guaranteed, since Bob does not necessarily couple $|\psi\rangle$ to 
the graph state, and Alice cannot check the correctness of the input state.
Bob might discard $|\psi\rangle$ and entangles 
completely different state $|\psi'\rangle$  
to the graph state. In this case what Alice obtains
is not $U|\psi\rangle$ but $U|\psi'\rangle$.
Can Alice verify that Bob honestly coupled her
input state to the graph state?

In this paper, we introduce a new protocol of
measurement-only verifiable blind quantum computing where
not only the computation itself but also
the quantum input are verifiable.
Our strategy is to combine the trap technique and
the stabilizer test. The correctness of the computing
is verified by the stabilizer test, and the correctness of the
quantum input is verified by checking the trap qubits that are
randomly hidden in the input state. When the traps are checked,
the state can be isolated from the graph state by
measuring the connecting qubits in $Z$ basis.
The main technical challenge in our proof is to show that
the trap verification and the stabilizer verification can 
coexist with each other.

\section{Stabilizer test}
We first review the stabilizer test.
Let us consider an $N$-qubit state $\rho$ and
a set $g\equiv\{g_1,...,g_n\}$ of generators of a stabilizer group.
The stabilizer test is a following test:
\begin{itemize}
\item[1.]
Randomly generate an $n$-bit string
$k\equiv(k_1,...,k_n)\in\{0,1\}^n$. 
\item[2.]
Measure the operator
\begin{eqnarray*}
s_k\equiv\prod_{j=1}^ng_j^{k_j}.
\end{eqnarray*}
Note that this measurement can be done with single-qubit
measurements, since $s_k$ is a tensor product of Pauli operators.
\item[3.]
If the result is $+1$ ($-1$), the test passes (fails). 
\end{itemize}
The probability of passing the stabilizer test is
\begin{eqnarray*}
p_{pass}=\frac{1}{2^n}\sum_{k\in\{0,1\}^n}\mbox{Tr}
\Big(\frac{I+s_k}{2}\rho\Big).
\end{eqnarray*}
We can show that if the probability of passing
the test is high, $p_{pass}\ge1-\epsilon$, then
$\rho$ is ``close" to a certain stabilized state $\sigma$ in the sense of
\begin{eqnarray}
\mbox{Tr}(M\sigma)(1-2\epsilon)-\sqrt{2\epsilon}
\le
\mbox{Tr}(M\rho)\le\mbox{Tr}(M\sigma)+\sqrt{2\epsilon}
\label{st}
\end{eqnarray}
for any POVM element $M$.

In fact,
if $p_{pass}\ge1-\epsilon$, we obtain
\begin{eqnarray*}
\mbox{Tr}\Big(\prod_{j=1}^k\frac{I+g_j}{2}\rho\Big)\ge1-2\epsilon.
\end{eqnarray*}
Let
\begin{eqnarray*}
	\Lambda\equiv
\prod_{j=1}^k\frac{I+g_j}{2}.
\end{eqnarray*}
From the gentle measurement lemma~\cite{gentle},
\begin{eqnarray*}
\frac{1}{2}\|\rho-\Lambda \rho\Lambda\|_1&\le&
\sqrt{1-\mbox{Tr}(\Lambda\rho)}\\
&\le&\sqrt{1-(1-2\epsilon)}\\
&=&\sqrt{2\epsilon}.
\end{eqnarray*}
Note that 
\begin{eqnarray*}
g_j\frac{\Lambda\rho\Lambda}{\mbox{Tr}(\Lambda\rho)}g_j
=
\frac{\Lambda\rho\Lambda}{\mbox{Tr}(\Lambda\rho)}
\end{eqnarray*}
for any $j$, and therefore, 
$\Lambda\rho\Lambda/\mbox{Tr}(\Lambda\rho)$ 
is a stabilized state.

For any positive operator $M$,
\begin{eqnarray*}
\mbox{Tr}(M\rho)-\mbox{Tr}(\Lambda\rho\Lambda)\le\sqrt{2\epsilon},
\end{eqnarray*}
which means
\begin{eqnarray*}
\mbox{Tr}(M\rho)&\le&
\mbox{Tr}\Big(M\frac{\Lambda\rho\Lambda}
{\mbox{Tr}(\Lambda\rho)}\Big)
{\mbox{Tr}(\Lambda\rho)}
+\sqrt{2\epsilon}\\
&\le&
\mbox{Tr}\Big(M\frac{\Lambda\rho\Lambda}
{\mbox{Tr}(\Lambda\rho)}\Big)
+\sqrt{2\epsilon}.
\end{eqnarray*}
And, for any positive operator $M$,
\begin{eqnarray*}
\mbox{Tr}(\Lambda\rho\Lambda)-\mbox{Tr}(M\rho)
\le\sqrt{2\epsilon},
\end{eqnarray*}
which means
\begin{eqnarray*}
\mbox{Tr}(M\rho)
&\ge&
\mbox{Tr}\Big(M\frac{\Lambda\rho\Lambda}{\mbox{Tr}(\Lambda\rho)}\Big)
\mbox{Tr}(\Lambda\rho)
-\sqrt{2\epsilon}\\
&\ge&
\mbox{Tr}\Big(M\frac{\Lambda\rho\Lambda}{\mbox{Tr}(\Lambda\rho)}\Big)
(1-2\epsilon)
-\sqrt{2\epsilon}.
\end{eqnarray*}

\section{Our protocol}
Now we explain our protocol and analyze it,
which is the main result of the present paper.
Let us consider the following situation:
Alice possesses an $m$-qubit state $|\psi\rangle$,
but she neither knows its classical description nor
a classical description of a quantum circuit that efficiently
generates the state.
(For example, she just receives $|\psi\rangle$ from
her friend Charlie, etc.) 
She wants to perform a polynomial-size quantum computing $U$
on the input $|\psi\rangle$, but she cannot do it by herself.
She therefore asks Bob, who is very powerful but not trusted, 
to perform her quantum computing.
We show that Alice can delegate her quantum computing to Bob without
revealing $|\psi\rangle$ and $U$,
and she can verify the correctness of the
computation and input.

For simplicity, we assume that Alice wants to solve a decision problem $L$.
Alice measures the output qubit of $U|\psi\rangle$ in the computational
basis, and accepts (rejects) if the result is 1 (0).
As usual, we assume that
for any yes instance $x$, i.e., $x\in L$, the acceptance probability
is larger than $a$, and for any no instance $x$, i.e., $x\notin L$,
the acceptance probability is smaller than $b$,
where $a-b\ge1/poly(|x|)$.

\begin{figure}[htbp]
\begin{center}
\includegraphics[width=0.2\textwidth]{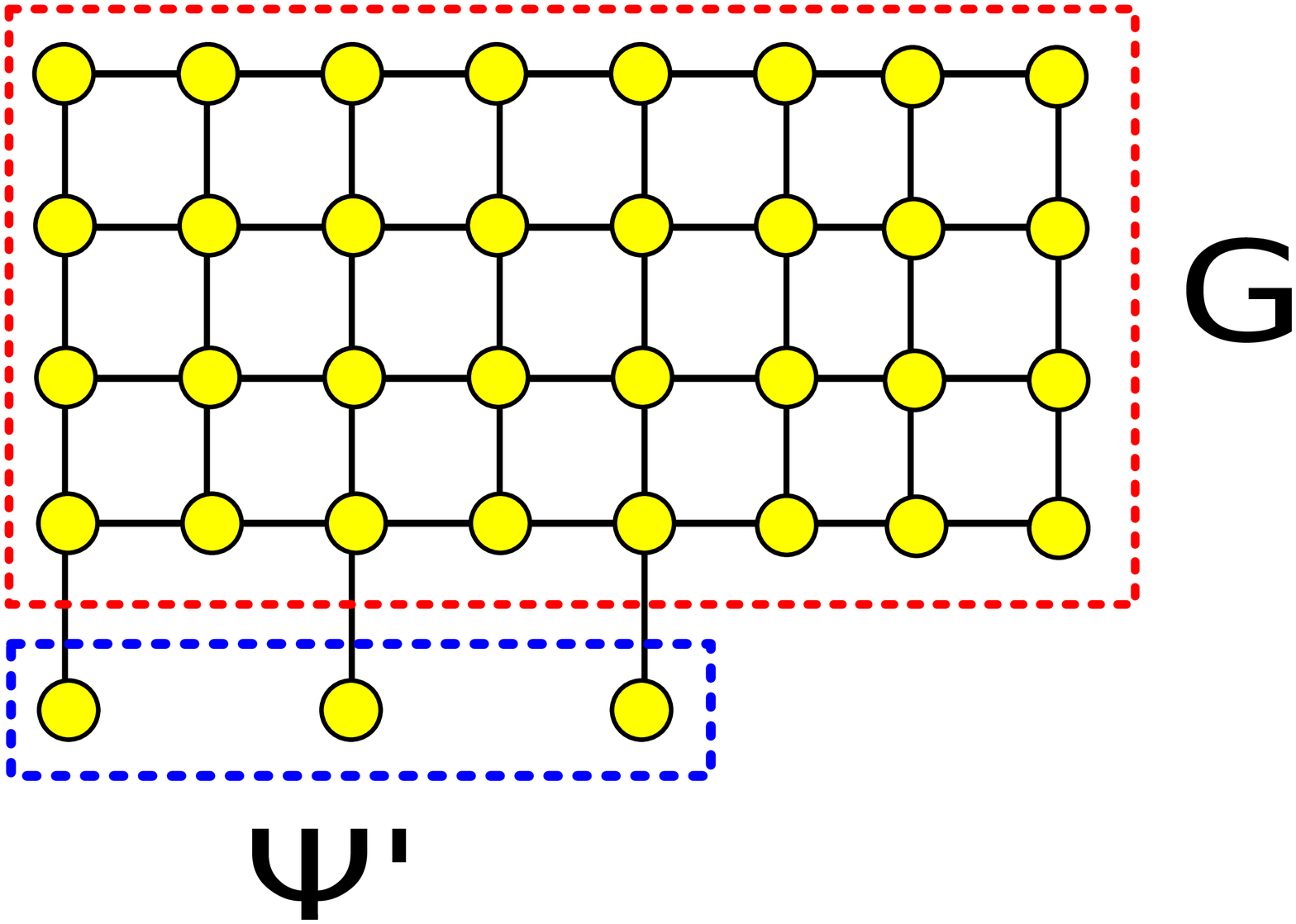}
\end{center}
\caption{
The state $|G_{\Psi'}\rangle$.
The state in the red dotted box is $|G\rangle$
and that in the blue dotted box is $|\Psi'\rangle$.
$E_{connect}$ is the set of edges that connects qubits
in the red dotted box and those in the blue dotted box. 
} 
\label{fig}
\end{figure}

Our protocol runs as follows:
\begin{itemize}
\item[1.]
Alice randomly chooses a $3m$-qubit permutation $P$, and applies it
on 
\begin{eqnarray*}
|\Psi\rangle\equiv
|\psi\rangle\otimes|0\rangle^{\otimes m}\otimes|+\rangle^{\otimes m}
\end{eqnarray*}
to generate $P|\Psi\rangle$.
Alice further chooses a random $6m$-bit string
$(x_1,...,x_{3m},z_1,...,z_{3m})\in\{0,1\}^{6m}$, and applies
$\bigotimes_{j=1}^{3m} X_j^{x_j}Z_j^{z_j}$
on $P|\Psi\rangle$ to generate
\begin{eqnarray*}
|\Psi'\rangle\equiv
\Big(\bigotimes_{j=1}^{3m} X_j^{x_j}Z_j^{z_j}\Big)P|\Psi\rangle.
\end{eqnarray*}
She sends $|\Psi'\rangle$ to Bob.
(Or, it is reasonable
to assume that Charlie gives Alice $|\Psi'\rangle$ and
information of $P$ and $(x_1,...,x_{3m},z_1,...,z_{3m})$
in stead of giving $|\psi\rangle$.)
\item[2.]
If Bob is honest, he generates the $(3m+N)$-qubit state
\begin{eqnarray}
|G_{\Psi'}\rangle\equiv 
\Big(\bigotimes_{e\in E_{connect}}CZ_e\Big)(|\Psi'\rangle\otimes|G\rangle),
\label{honestBob}
\end{eqnarray}
and sends each qubit of it one by one to Alice,
where $CZ_e$ is the CZ gate on the vertices of the edge $e$,
and $E_{connect}$ is the set of edges that connects qubits
in $|G\rangle$ and $|\Psi'\rangle$ (Fig.~\ref{fig}).
If Bob is malicious, he sends any $(3m+N)$-qubit state $\rho$
to Alice.
\item[3.]
\begin{itemize}
\item[3-a.]
With probability $q$, which is specified later, Alice does
the measurement-based quantum computing on qubits sent from Bob.
If the computation result is accept (reject), she accepts (rejects).
(During the computation, Alice of course corrects the
initial random Pauli operator 
and permutation
$(\bigotimes_{j=1}^{3m}X_j^{x_j}Z_j^{z_j})P$.) 
\item[3-b.]
With probability $(1-q)/2$, Alice does the stabilizer test,
and if she passes (fails) the test, she accepts (rejects).
\item[3-c.]
With probability $(1-q)/2$, Alice does the following
test, which we call the input-state test:
Let $V_1$ and $V_2$ be the set of qubits in the red dotted box
and the blue dotted box in Fig.~\ref{fig}, respectively.
Alice stores qubits in $V_2$
in her memory, and measures each qubit in $V_1$ in
$Z$ basis. 
If the $Z$-basis measurement result on the 
nearest-neighbour of $j$th vertex in $V_2$ is 1,
Alice applies $Z$ on the $j$th vertex in $V_2$
for $j=1,...,3m$.
If Bob was honest, the state of $V_2$ is now
\begin{eqnarray*}
|\Psi'\rangle=\Big(\bigotimes_{j=1}^{3m}X_j^{x_j}Z_j^{z_j}\Big)P|\Psi\rangle.
\end{eqnarray*}
Alice further applies 
$P^\dagger(\bigotimes_{j=1}^{3m} X_j^{x_j}Z_j^{z_j})$ on $V_2$.
If Bob was honest, the state of $V_2$ is now
$|\Psi\rangle=
|\psi\rangle\otimes|0\rangle^{\otimes m}\otimes|+\rangle^{\otimes m}$.
Then Alice performs the projection measurement
$
\{\Lambda_0\equiv|0\rangle\langle0|^{\otimes m}\otimes
|+\rangle\langle+|^{\otimes m},
\Lambda_1=I^{\otimes 2m}-\Lambda_0\}$,
on the last $2m$ qubits of $V_2$.
If she gets $\Lambda_0$, she accepts.
Otherwise, she rejects.
\end{itemize}
\end{itemize}

Now let us analyze the protocol.
First, the blindness is obvious, since what she sends to
Bob is the completely-mixed state from Bob's view point, 
and due to the no-signaling principle,
Alice's operations on states sent from Bob
do not transmit any information to Bob.

Second, let us consider the case of $x\in L$.
In this case, honest Bob generates the correct state, 
$|G_{\Psi'}\rangle$,
Eq.~(\ref{honestBob}),
and therefore Alice can do correct computation,
if she chooses the measurement-based quantum computing,
passes the stabilizer test with probability 1,
if she chooses the stabilizer test,
and passes the input-state test with probability 1,
if she chooses the input-state test.
Therefore, the acceptance probability, $p_{acc}^{x\in L}$, is
\begin{eqnarray*}
p_{acc}^{x\in L}
&\ge& qa+\frac{1-q}{2}\times 1+\frac{1-q}{2}\times 1\\
&=&qa+(1-q)\equiv\alpha.
\end{eqnarray*}

Finally, let us consider the case of $x\notin L$. 
In this case, Bob might be malicious, and
can send any $(3m+N)$-qubit state $\rho$.
Let $p_{Gpass}$ and $p_{\psi pass}$ be the probability
of passing the stabilizer test and the initial-state test,
respectively.

Let $\epsilon=\frac{1}{poly(|x|)}$.
It is easy to see that the acceptance probability, $p_{acc}^{x\notin L}$, is
given as follows:
\begin{itemize}
\item[1.]
If $p_{Gpass}\ge 1-\epsilon$ and $p_{\psi pass}<1-\epsilon$, then
\begin{eqnarray*}
p_{acc}^{x\notin L}&<&q+\frac{1-q}{2}+\frac{1-q}{2}(1-\epsilon)\equiv\beta_1.
\end{eqnarray*}
\item[2.]
If $p_{Gpass}< 1-\epsilon$ and $p_{\psi pass}\ge1-\epsilon$, then
\begin{eqnarray*}
p_{acc}^{x\notin L}&<&q+\frac{1-q}{2}(1-\epsilon)+\frac{1-q}{2}=\beta_1.
\end{eqnarray*}
\item[3.]
If $p_{Gpass}< 1-\epsilon$ and $p_{\psi pass}<1-\epsilon$, then
\begin{eqnarray*}
p_{acc}^{x\notin L}&<&q
+\frac{1-q}{2}(1-\epsilon)
+\frac{1-q}{2}(1-\epsilon)\\
&=&
q+(1-q)(1-\epsilon)
\equiv\beta_2.
\end{eqnarray*}
\end{itemize}

Let us consider the remaining case,
$p_{Gpass}\ge 1-\epsilon$ and $p_{\psi pass}\ge1-\epsilon$. 
From the triangle inequality and the invariance of the trace norm
under a unitary operation,
we obtain
\begin{eqnarray*}
	\frac{1}{2}\big\|\rho-G_{\Psi'}\big\|_1
&=&
\frac{1}{2}\big\|
\rho-G_\sigma+G_\sigma-G_{\Psi'}\big\|_1\\
&\le&
\frac{1}{2}\big\|\rho-G_\sigma\big\|_1
+\frac{1}{2}\big\|G_\sigma-G_{\Psi'}\big\|_1\\
&=&
\frac{1}{2}\big\|\rho-G_\sigma\big\|_1
+\frac{1}{2}\big\|\sigma-|\Psi'\rangle\langle\Psi'|\big\|_1,
\end{eqnarray*}
where $G_{\Psi'}\equiv|G_{\Psi'}\rangle\langle G_{\Psi'}|$,
\begin{eqnarray*}
G_\sigma\equiv 
\big(\bigotimes_{e\in E_{connect}}CZ_e\big)
(\sigma\otimes|G\rangle\langle G|)
\big(\bigotimes_{e\in E_{connect}}CZ_e\big),
\end{eqnarray*}
and $\sigma$ is any $3m$-qubit state on $V_2$.
Since $p_{Gpass}\ge1-\epsilon$, the first term is upperbounded
as
\begin{eqnarray*}
\frac{1}{2}\big\|\rho-G_\sigma\big\|_1\le\sqrt{2\epsilon},
\end{eqnarray*}
from Eq.~(\ref{st}).
We can show that if $p_{\psi pass}\ge1-\epsilon$,
the second term is upperbounded as
\begin{eqnarray}
\frac{1}{2}\big\|\sigma-|\Psi'\rangle\langle\Psi'|\big\|_1
\le\sqrt{2\epsilon}+\sqrt{\frac{2}{3}+\epsilon}.
\label{psipass}
\end{eqnarray}
The proof is given in Appendix.

Therefore, we obtain
\begin{eqnarray*}
\frac{1}{2}\big\|\rho-G_{\Psi'}\big\|_1
&\le&2\sqrt{2\epsilon}+\sqrt{\frac{2}{3}+\epsilon}\\
&\equiv&\delta,
\end{eqnarray*}
which means
\begin{eqnarray*}
\big|\mbox{Tr}(\Pi \rho)-\mbox{Tr}(\Pi G_{\Psi'})\big|
\le\delta
\end{eqnarray*}
for the POVM element $\Pi$ corresponding to the acceptance
of the measurement-based quantum computing.
Therefore, the total acceptance probability, $p_{acc}^{x\notin L}$, is
\begin{eqnarray*}
p_{acc}^{x\notin L}&\le& q(b+\delta)+\frac{1-q}{2}+\frac{1-q}{2}\\
&=&q(b+\delta)+(1-q)\equiv\beta_3.
\end{eqnarray*}

Let us define
\begin{eqnarray*}
\Delta_1(q)&\equiv&\alpha-\beta_1=q(a-1)+\frac{\epsilon(1-q)}{2},\\
\Delta_2(q)&\equiv&\alpha-\beta_2=q(a-1)+\epsilon (1-q),\\
\Delta_3(q)&\equiv&\alpha-\beta_3=q(a-b-\delta).
\end{eqnarray*}
The optimal value 
\begin{eqnarray*}
q^*\equiv\frac{\frac{\epsilon}{2}}{1+\frac{\epsilon}{2}-b-\delta}
\end{eqnarray*}
of $q$ is that satisfies $\Delta_1(q)=\Delta_3(q)$.
Then, if we take $a=1-2^{-r}$ and $b=2^{-r}$ for a polynomial $r$,
\begin{eqnarray*}
\Delta_3(q^*)&=&
\frac{\frac{\epsilon}{2}(a-b-\delta)}{1+\frac{\epsilon}{2}-b-\delta}\\
&\ge&
\frac{\epsilon}{4}\Big(1-2^{-r+1}-2\sqrt{2\epsilon}
-\sqrt{\frac{2}{3}+\epsilon}\Big)\\
&\ge& \frac{1}{poly(|x|)}.
\end{eqnarray*}
As usual, the inverse polynomial gap can be amplified with a polynomial
overhead. 

\acknowledgements
TM is supported by the
Grant-in-Aid for Scientific Research on Innovative Areas
No.15H00850 of MEXT Japan, and the Grant-in-Aid
for Young Scientists (B) No.26730003 of JSPS.

\appendix*
\begin{widetext}
\section{Proof of Eq.~(\ref{psipass})}
In this appendix, we show Eq.~(\ref{psipass}).
Due to the triangle inequality and the invariance of the trace norm
under a unitary operation,
\begin{eqnarray*}
\frac{1}{2}\big\|\sigma-|\Psi'\rangle\langle\Psi'|\big\|_1
&=&
\frac{1}{2}\Big\|\sigma-\big(\bigotimes_{j=1}^{3m}X_j^{x_j}Z_j^{z_j}\big)
P\big(\fbox{$\psi$}\otimes\fbox{0}^{\otimes m}\otimes\fbox{+}^{\otimes m}\big)
P^\dagger
\big(\bigotimes_{j=1}^{3m}X_j^{x_j}Z_j^{z_j}\big)
\Big\|_1\\
&=&
\frac{1}{2}\Big\|
P^\dagger
\big(\bigotimes_{j=1}^{3m}X_j^{x_j}Z_j^{z_j}\big)
\sigma
\big(\bigotimes_{j=1}^{3m}X_j^{x_j}Z_j^{z_j}\big)
P
-
\fbox{$\psi$}\otimes\fbox{0}^{\otimes m}\otimes\fbox{+}^{\otimes m}
\Big\|_1\\
&\le&
\frac{1}{2}\Big\|
P^\dagger
\big(\bigotimes_{j=1}^{3m}X_j^{x_j}Z_j^{z_j}\big)
\sigma
\big(\bigotimes_{j=1}^{3m}X_j^{x_j}Z_j^{z_j}\big)
P
-\rho_{before}\Big\|_1\\
&&+\frac{1}{2}\Big\|
\rho_{before}-
\fbox{$\psi$}\otimes\fbox{0}^{\otimes m}\otimes\fbox{+}^{\otimes m}
\Big\|_1,
\end{eqnarray*}
where
$\rho_{before}$ is  the state
before measuring $\{\Lambda_0,\Lambda_1\}$.

From the monotonicity of the trace distance under
a CPTP map, the first term is upperbounded as
\begin{eqnarray*}
\frac{1}{2}\Big\|
P^\dagger
\big(\bigotimes_{j=1}^{3m}X_j^{x_j}Z_j^{z_j}\big)
\sigma
\big(\bigotimes_{j=1}^{3m}X_j^{x_j}Z_j^{z_j}\big)
P
-\rho_{before}\Big\|_1
\le
\frac{1}{2}\|G_\sigma-\rho\|_1\le\sqrt{2\epsilon}.
\end{eqnarray*}

As is shown below,
the second term is upperbounded as
\begin{eqnarray}
\frac{1}{2}\Big\|~
\fbox{$\psi$}\otimes\fbox{0}^{\otimes m}\otimes\fbox{+}^{\otimes m}
-\rho_{before}\Big\|_1
&\le&\sqrt{\frac{2}{3}+\epsilon}.
\label{secondterm}
\end{eqnarray}
Therefore, we have shown Eq.~(\ref{psipass}).

Let us show Eq.~(\ref{secondterm}).
Note that
\begin{eqnarray*}
\rho_{before}=
\frac{1}{(3m)!}\frac{1}{4^{3m}}
\sum_{P,\alpha,k}
P^\dagger\sigma_\alpha
E_k\sigma_\alpha 
P\big(
\fbox{$\psi$}\otimes \fbox{0}^{\otimes m}\otimes \fbox{+}^{\otimes m}\big)
P^\dagger
\sigma_\alpha E_k^\dagger
\sigma_\alpha
P,
\end{eqnarray*}
where $\sigma_\alpha$ is a $3m$-qubit Pauli operator,
and $E_k$ is a Kraus operator.
Let us decompose each Kraus operator in terms
of Pauli operators as $E_k=\sum_{\beta}C_\beta^k\sigma_\beta$.
Since
\begin{eqnarray*}
I&=&\sum_kE_k^\dagger E_k\\
&=&\sum_{k,\beta,\gamma}C_\beta^{k*}C_\gamma^k
\sigma_\beta\sigma_\gamma\\
&=&
\sum_{k,\beta}|C_\beta^k|^2I
+\sum_{k,\beta\neq\gamma}C_\beta^{k*}C_\gamma^k
\sigma_\beta\sigma_\gamma,
\end{eqnarray*}
we obtain
\begin{eqnarray*}
\sum_{k,\beta}|C_\beta^k|^2=1.
\end{eqnarray*}

Then,
\begin{eqnarray*}
\rho_{before}&=&\frac{1}{(3m)!}\frac{1}{4^{3m}}
\sum_{P,\alpha,k,\beta,\gamma}
C_\beta^kC_\gamma^{k*}
P^\dagger\sigma_\alpha
\sigma_\beta\sigma_\alpha 
P\big(\fbox{$\psi$}
\otimes \fbox{0}^{\otimes m}\otimes \fbox{+}^{\otimes m}\big)P^\dagger
\sigma_\alpha \sigma_\gamma
\sigma_\alpha P\\
&=&
\frac{1}{(3m)!}
\sum_{P,k,\beta}
|C_\beta^k|^2
P^\dagger
\sigma_\beta
P\big(\fbox{$\psi$}
\otimes \fbox{0}^{\otimes m}\otimes \fbox{+}^{\otimes m}\big)P^\dagger
\sigma_\beta P\\
&=&
\frac{1}{(3m)!}
\sum_{P,\beta}
D_\beta
P^\dagger
\sigma_\beta
P\big(\fbox{$\psi$}\otimes 
\fbox{0}^{\otimes m}\otimes \fbox{+}^{\otimes m}\big)P^\dagger
\sigma_\beta P\\
&=&
D_0
\big(\fbox{$\psi$}\otimes \fbox{0}^{\otimes m}\otimes \fbox{+}^{\otimes m}\big)
+
\frac{1}{(3m)!}
\sum_{P,\beta\neq0}
D_\beta
P^\dagger
\sigma_\beta
P\big(\fbox{$\psi$}
\otimes \fbox{0}^{\otimes m}\otimes \fbox{+}^{\otimes m}\big)P^\dagger
\sigma_\beta P\\
&\equiv&\rho_1+\rho_2.
\end{eqnarray*}
Here, $\sigma_0=I^{\otimes 3m}$, we have used the relation
\begin{eqnarray*}
\sum_{\alpha}\sigma_\alpha\sigma_\beta\sigma_\alpha
\rho\sigma_\alpha\sigma_\gamma\sigma_\alpha
=0
\end{eqnarray*}
for any $\rho$ and $\beta\neq\gamma$,
and defined
\begin{eqnarray*}
\sum_k|C_\beta^k|^2=D_\beta.
\end{eqnarray*}
Note that
\begin{eqnarray*}
\sum_\beta D_\beta=\sum_{\beta,k}|C_\beta^k|^2=1.
\end{eqnarray*}

It is obvious that
\begin{eqnarray*}
\mbox{Tr}\Big[
\big(I^{\otimes m}-\fbox{$\psi$}\big)\otimes\fbox{0}^{\otimes m}
\otimes \fbox{+}^{\otimes m}
\times
\rho_1
\Big]=0.
\end{eqnarray*}
Furthermore,
\begin{eqnarray*}
&&\mbox{Tr}\Big[
\big(I^{\otimes m}-\fbox{$\psi$}\big)
\otimes\fbox{0}^{\otimes m}\otimes \fbox{+}^{\otimes m}
\times
\rho_2
\Big]\\
&=&
\frac{1}{(3m)!}
\sum_{P,\beta\neq0}
D_\beta
\mbox{Tr}\Big[
\big(I^{\otimes m}-\fbox{$\psi$}\big)\otimes
\fbox{0}^{\otimes m}\otimes \fbox{+}^{\otimes m}
\times
P^\dagger
\sigma_\beta
P\big(\fbox{$\psi$}\otimes \fbox{0}^{\otimes m}\otimes 
\fbox{+}^{\otimes m}\big)P^\dagger
\sigma_\beta P
\Big]\\
&\le&
\frac{1}{(3m)!}
\sum_{\beta\neq0}
D_\beta
(2m\times(3m-1)!)\\
&\le&
\frac{2m\times(3m-1)!}{(3m)!}\\
&=&\frac{2}{3}.
\end{eqnarray*}

Therefore,
\begin{eqnarray*}
\mbox{Tr}\Big[
\big(I^{\otimes m}-\fbox{$\psi$}\big)
\otimes\fbox{0}^{\otimes m}\otimes \fbox{+}^{\otimes m}
\times
\rho_{before}
\Big]\le\frac{2}{3},
\end{eqnarray*}
which means
\begin{eqnarray*}
\mbox{Tr}\Big[
\fbox{$\psi$}\otimes\fbox{0}^{\otimes m}\otimes \fbox{+}^{\otimes m}
\times
\rho_{before}
\Big]
&\ge&
\mbox{Tr}\Big[
I^{\otimes m}\otimes\fbox{0}^{\otimes m}\otimes \fbox{+}^{\otimes m}
\times
\rho_{before}
\Big]-\frac{2}{3}\\
&\ge&
1-\epsilon-\frac{2}{3}\\
&=&
\frac{1}{3}-\epsilon,
\end{eqnarray*}
where we have used the assumption that
\begin{eqnarray*}
p_{\psi pass}=
\mbox{Tr}\Big[
I^{\otimes m}\otimes\fbox{0}^{\otimes m}\otimes \fbox{+}^{\otimes m}
\times
\rho_{before}\Big]\ge1-\epsilon.
\end{eqnarray*}
Therefore
\begin{eqnarray*}
\frac{1}{2}\Big\|~
\fbox{$\psi$}\otimes\fbox{0}^{\otimes m}\otimes\fbox{+}^{\otimes m}
-\rho_{before}\Big\|_1
&\le&\sqrt{1-\Big(\frac{1}{3}-\epsilon\Big)}\\
&=&\sqrt{\frac{2}{3}+\epsilon}.
\end{eqnarray*}

\end{widetext}

\end{document}